\documentclass[%
 reprint,
 prd,
 amsmath,amssymb,
 aps,
 lengthcheck,
floatfix,
]{revtex4}

\usepackage{graphicx}
\usepackage{dcolumn}
\usepackage{bm}
\usepackage[colorlinks=true,linktocpage=true,linkcolor=blue,citecolor=blue,allcolors=
blue] {hyperref }
\usepackage[utf8]{inputenc}
\usepackage{multirow}
\usepackage{soul}
\usepackage{float}
\usepackage{graphicx}
\usepackage{times}
\usepackage[normalem]{ulem}
\usepackage{color}
\usepackage{cellspace}

\usepackage{lipsum}

\newcommand{\bef}{\begin{figure}}
\newcommand{\eef}{\end{figure}}
\newcommand{\bc}{\begin{center}}
\newcommand{\ec}{\end{center}}

\newcommand{\antinue}{\ensuremath{\overline{\nu}_{e}}}

\newcommand{\comment}[1]{}
\renewcommand\sout{\bgroup \color{red} \ULdepth=-.5ex \ULset}

\begin{document}


\title{Sterile neutrino searches with reactor antineutrinos using coherent 
neutrino-nucleus scattering experiments}

%

\author{S.P. Behera$^{1}$}
\email{shiba@barc.gov.in}
\author{D.K. Mishra$^{1}$}
\author{P.K. Netrakanti$^{1}$}%
\author{R.Sehgal$^{1}$}%
\author{Kirtikesh Kumar$^{1}$}%
\author{R.Dey$^{1}$$^{,2}$}%
\author{V. Jha$^{1}$$^{,2}$}%
\affiliation{$^{1}$Nuclear Physics Division, Bhabha Atomic Research Centre,\\ Mumbai - 400085, India}
\affiliation{$^{2}$Homi Bhabha National Institute, Anushakti Nagar, Mumbai - 400094, India}

\begin{abstract}
We present an analysis on the sensitivity to the active-sterile neutrino mixing with 
Germanium (Ge) and Silicon (Si) detectors in the context of the proposed coherent 
elastic neutrino-nucleus experiment in India. The study has been carried out with 3 
(active) $+$ 1 (sterile) neutrino oscillation model. It is observed that the 
measurements that can be carried out with the Ge detector exhibit better 
sensitivity to the active-sterile neutrino mixing as compared to the Si detector. Both  
detectors are able to exclude most of the anomaly regions observed by the GALLIUM 
experiment. The Ge detector with mass 10 kg, can observe the active-sterile 
neutrino oscillation at 95$\%$ confidence level, provided that
$\sin^{2}2\theta_{14}\geq 0.09$ at $\Delta m^{2}_{41}$ = 1.0 eV$^{2}$ for an
exposure of 1-yr. At higher values of $\Delta m^{2}_{41}$, a better sensitivity is 
obtained at a short baseline. It is also found that the threshold as well as 
resolution of the detectors play a crucial role on the measurements of 
active-sterile neutrino mixing parameters.
\end{abstract}


\maketitle
\section{Introduction}
\label{sec:intro}
The concept of neutrino was first introduced by Pauli in 1930 while explaining the 
energy spectra of beta particles. Later on, it was first observed by Cowan and Reines 
via the inverse beta decay (IBD) process using the reactor as a source. The small 
mass of neutrinos results from the combination of three mass eigen-states, which is
established by many 
experiments using solar, atmospheric, reactor, and accelerator-based neutrinos. 
However, the physical origin of their masses is still not understood. Since, 
neutrinos do not have fixed mass but are described by quantum mechanical 
superposition mass eigen-states ( $\nu_{1}$, $\nu_{2}$, $\nu_{3}$), with each one 
having distinct mass eigen-value $m_{1}$, $m_{2}$, $m_{3}$.Neutrinos can change flavor 
while moving from one place to another, a phenomenon commonly referred to as neutrino 
flavor oscillation. At 
present several efforts are going on for the precise determination of neutrino 
oscillation parameters~\cite{JUNO:2015zny,DUNE:2020jqi}. Many experiments
are being performed to obtain more accurate values 
of neutrino-nucleus cross-sections. Coherent elastic neutrino-nucleus scattering 
(CE$\nu$NS) is a standard model (SM) process where the low energy neutrinos scatter 
off the atomic nucleus coherently through the neutral-current weak 
interactions~\cite{Freedman:1973yd}. For low-energy neutrinos ($<$ 50 MeV), the 
CE$\nu$NS process has a larger cross-section for neutron-rich targets compared to 
other known processes, such as inverse beta decay (IBD) and neutrino-electron 
scattering~\cite{CONUS:2021dwh}, which have been traditionally used for neutrino 
measurements. Further, the CE$\nu$NS is a threshold-less process in contrast to IBD. 
Although, having a larger cross-section, the CE$\nu$NS process has not been observed 
earlier due to the challenge of measuring the low-energy recoil nuclei. The 
measurement of CE$\nu$NS cross-section requires a high flux of low energy neutrinos 
and measurement of low nuclear recoil energies.

There are mainly three possible sources of (anti)neutrinos that could be used to 
search for the measurement of the CE$\nu$NS process, namely pion decay at rest (DAR)
beam, an intense 
radioactive source, or a nuclear reactor. Recently, the CE$\nu$NS cross-section has 
been measured by the COHERENT experiment using the neutrinos produced from the 
spallation neutron source as a DAR~\cite{COHERENT:2017ipa}. The neutrinos produced 
due to this method have a maximum energy of about 53 MeV. The endeavor to 
measure CE$\nu$NS using neutrinos from a nuclear reactor or an intense radioactive 
source presents both challenges and intriguing possibilities. In the case of 
reactor, \antinue s produced due to nuclear fission have an endpoint energy 
near 10 MeV. Due to the lower neutrino energies as compared to other sources, it 
becomes imperative to employ novel detector technologies capable of achieving 
detection thresholds in the range of a few tens of eV. Various cryogenic bolometers 
stand out as promising candidates for optimization to meet this need.

The measurement of CE$\nu$NS cross-section by the COHERENT group has opened up an 
avenue for exploring various aspects related to physics beyond the SM at low 
energy. At present, several experiments are going on and some are proposed to 
measure the CE$\nu$NS cross-section with the required threshold by using charge-coupled 
devices (CCDs), metallic superconducting bolometers, Ge-based semiconductor 
detectors, and the Scintillating Bubble Chamber~\cite{Alfonso-Pita:2022eli}
employing reactor \antinue~ as a 
source~\cite{Colaresi:2022obx,MINER:2016igy,Akimov:2017hee,Strauss:2017cuu,CONUS:2021dwh,
  CONNIE:2019swq,CONNIE:2021ggh,Fernandez-Moroni:2020yyl}.
The measurement of the CE$\nu$NS process can shed light on several fundamental SM 
physics aspects such as non-standard interactions~\cite{Coloma:2017ncl,Liao:2017uzy},
neutron density distribution~\cite{Cadeddu:2017etk}, neutrino magnetic
moment~\cite{Kosmas:2015sqa}, and the weak mixing 
angle~\cite{Canas:2018rng,Papoulias:2017qdn}. The CE$\nu$NS is a flavor-blind 
process, hence flavor-independent astronomy with supernova neutrinos becomes 
feasible, which allows us to investigate the interior of dense objects as well as 
stellar evolution in details~\cite{Biassoni:2011xuo,Brdar:2018zds}.

In this context, we propose to measure the CE$\nu$NS process using the reactor 
\antinue s in India, to address various fundamental physics aspects of neutrino as mentioned earlier. The present study focused on investigating the possible explanation for an anomalous behavior found in several SBL reactor experiments that measure the $\antinue$s through
  the IBD process~\cite{Mention:2011rk}. 
The precise recalculation of the $\antinue$ flux by Mueller 
$et~al.$~\cite{Mueller:2011nm} and Huber~\cite{Huber:2011wv} shows about 6$\%$ deficit in the observed-to-predicted 
ratio of events at a short distance through 
the IBD process, which is known as the ``reactor antineutrino anomaly" 
(RAA)~\cite{Mention:2011rk}. There are two distinct explanations proposed  for this 
discrepancy. One of them is the disappearance of \antinue~while propagating from the 
source to detector due to Active-Sterile Neutrino (ASN) oscillations with the mass 
square difference $\Delta\mathrm{m^2}\sim$ 1 eV$^2$.
\begin{figure}
\advance\leftskip -50cm
\centering
\includegraphics[trim={3cm 0 10cm 0cm}, width=0.4\linewidth]{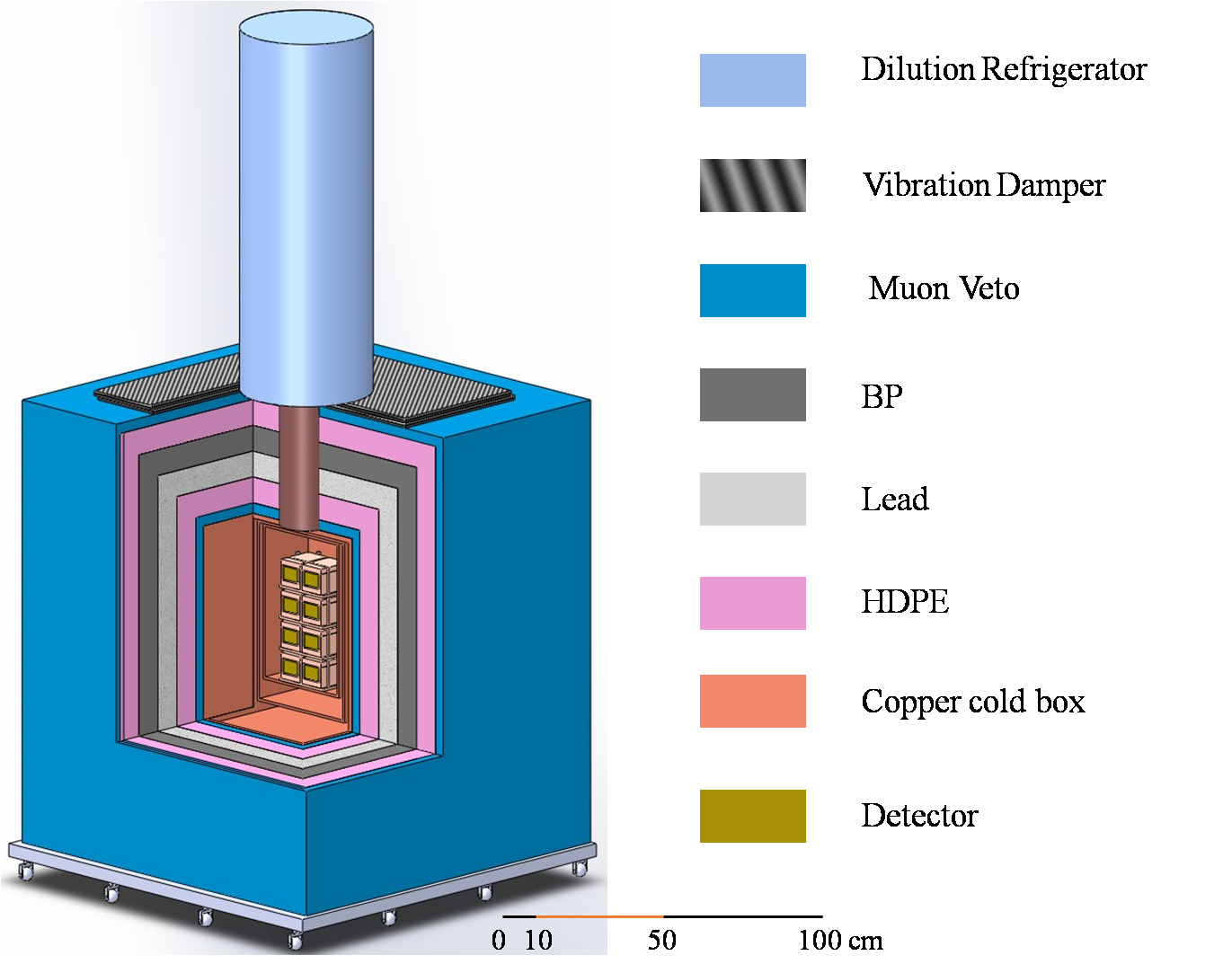}
\caption{ \label{fig:DetSetup} Schematic representation of experimental setup
  for CE$\nu$NS measurement. BP: boronated polyethylene, thickness: 10 cm, HDPE: high
  density polyethylene,thickness: 10 cm} 
\end{figure}
On the other hand, the observed discrepancy is likely to be related to the 
inaccuracies in predicting antineutrino flux, stemming from incomplete 
reactor models or uncertainties in nuclear data. The Huber-Muller
model utilizes the  cumulative $\beta^{-}$-spectra
measured at ILL for conversion to antineutrino 
spectra~\cite{Mueller:2011nm,Huber:2011wv}.
This suggests that experimental biases, possibly related to these measurements,
  might account for the anomaly.
Recent measurements indicate that bias in the prediction of $^{235}$U flux may be the 
likely cause of RAA~\cite{DayaBay:2017jkb,RENO:2018pwo}.
The CE$\nu$NS process has an advantage compared to other techniques for finding the 
possible existence of the sterile neutrino. It is a neutral current process in which 
neutrinos scatter off the nuclei, and are independent of the neutrino flavor. 
Therefore, any finding of an oscillation structure would signify complete mixing
with non-active neutrinos. 

Recently, various experiments have carried out the reactor
  \antinue s measurement  to study the ASN oscillations using the IBD process. The DANSS 
collaboration has measured the positron energy spectra at 3 different distances 
(ranging from 10.7 m to 12 m ) from the reactor core. From the measurements, a large 
fraction of the RAA region in the sin$^2 2\theta_{14} - \Delta m^{2}_{41}$ plane that 
covers the parameter space up to sin$^2 2\theta_{14} <$ 0.01~\cite{DANSS:2018fnn} are 
excluded. Similarly, the STEREO~\cite{STEREO:2018rfh} collaboration has measured the 
\antinue~ energy spectrum in six different detector cells covering baselines between 
9 and 11 meters from the reactor core of the ILL research reactor. The results based 
on the current reactor ON data are explained by the null ASN oscillation hypothesis 
and the best fit of the RAA can be excluded at 97.5$\%$ confidence level (C.L.). The 
reactor \antinue~ spectra measured by PROSPECT collaboration disfavors the RAA 
best-fit point at 2.2$\sigma$ C.L. and constrains significant portions of the 
previously allowed parameter space at 95$\%$ C.L.\cite{PROSPECT:2018dtt}. The 
Neutrino-4 group has measured \antinue~ energy spectra with the segmented detectors 
at different positions ranging from 6 to 12 meters. Their model-independent analysis 
excludes the RAA region at C.L. more than 3$\sigma$. However, the experiment has 
observed ASN oscillation at sin$^2 2\theta_{14}$ = 0.39 and 
$\Delta m^{2}_{41}$ = 7.3 $eV^2$ at C.L. of 2.8$\sigma$~\cite{NEUTRINO-4:2018huq}.
The above experimental observations are found by detecting \antinue
   through the IBD process. There are several experiments currently proposed or planned to
  investigate the mixing of active-sterile neutrinos using CE$\nu$NS
  process~\cite{MINER:2016igy,CONNIE:2016nav}. 
In pursuit of this goal, a feasibility study has been 
carried out to ascertain the ASN mixing sensitivity of various 
types of detectors by placing them at a short baseline (L$\leq$ 30m)
through CE$\nu$NS channel.

The article is organized as follows. In the subsequent section, a
comprehensive description  of the detector setup is provided. 
Antineutrinos production and various types of reactors used in this study are 
presented in Sec.\ref{sec:reactor}. The CE$\nu$NS process and the underlying 
detection principle are expounded upon in Sec.~\ref{sec:cens}.
An estimation of the expected number of events in the detector is presented
  in Section~\ref{sec:expect}. 
The phenomenon of ASN oscillation at short baseline considering the `3+1' 
mixing model is described in Sec.~\ref{sec:osciprob}. The simulation procedure for 
incorporation of detector response on coherent neutrino discussed in 
Sec.~\ref{sec:simul}. The sensitivity of the proposed experiment, statistical method 
 on $\chi^{2}$ estimation considered in this study is discussed in 
 Sec.~\ref{sec:chisq}. The sensitivity to sterile neutrino mixing at an exposure of
 one year is elaborated in Sec.~\ref{sec:results}. Finally, in Sec.~\ref{sec:summary},
 we summarize our observations and discuss the implication of this work.
\begin{table*}
 \begin{center}
\caption{\label{tab:reactortype}{Reactor details}}
\begin{tabular}{ ccc}
    Reactors name & Thermal power(MW$_{th}$) & Fuel type \\
\hline
   U-Apsara & 3.0 & U$_{3}$Si$_2$-Al (17$\%$  enriched $^{235}$U) \\
    DHRUVA & 100.0 & Natural uranium (0.7$\%$  $^{235}$U)  \\
 PFBR & 1250.0 & MOX(PuO$_{2}$-UO$_{2}$)  \\ 
 VVER & 3000.0 &  UO$_{2}$ (3.92 $\%$ enriched $^{235}$U)  \\
   \hline
\end{tabular}
\end{center}
\end{table*}
\section{Detector Setup at CE$\nu$NS Measurement }
\label{detSetup}
Figure~\ref{fig:DetSetup} shows the conceptual schematic of the detector setup 
proposed for the measurement of CE$\nu$NS using antineutrinos produced from the 
reactor. The core detector volume is covered with various shielding materials for 
reducing backgrounds.
The active volume of the detector is placed at the center within a cold copper box that
is surrounded by 4 cm thick plastic scintillator plates for vetoing cosmic muon.
High-density polyethylene(HDPE) sheets of thickness 10 cm are placed next
to the muon veto detectors for thermalizing the fast neutrons. A Lead shielding of 10 cm 
thickness is added to reduce the natural and reactor-generated 
gamma-rays background. Boronated polyethylene (BP) sheets with 15$\%$ boron content
and thickness of 10 cm are positioned at the outer part of the setup to attenuate
the thermal neutrons. The thickness of Lead and BP sheets are chosen considering the
previous study performed for another reactor-based antineutrino experiment using the ISMRAN
detector~\cite{Mulmule:2018efw}. To maintain the required low temperature for detector
  operation,  dilution refrigerator atop a vibration damper is placed on the outer
  layer of the shielding.
A link is established by a cold finger arrangement of length 50 cm between 
the refrigerator and detector, to maintain the temperature stability in
the active volume. The detectors will be housed within
a copper-cooled box of dimension 50 $\times$ 50 $\times$ 75 cm$^3$, with a
  2 cm thick copper plate.
The proposed detector will be placed at 4m distance from the movable U-Apsara
reactor core in the designated area within the reactor hall.

The incoming 
particle energy can be determined from particle interactions in a target which includes 
the measurement of ionization, scintillation, and/or the phonon excitation in
the material.
    The ionization signal is generated as the electron-hole pairs, which are produced during the
    energy loss process and collected on the electrodes. The metal electrodes on the two faces
of the crystal substrate acting as sensors to measure ionization. Simultaneously, a thermal sensor
in contact with the Si/Ge crystal records the phonon signal, similar to the one
discussed in ~\cite{CDMS:2005jsf}.
    Currently, two types of detectors such as high-purity Si or Ge are under consideration
    for measuring the signal produced by recoil nuclei. Each crystal is equipped with two
    concentric ionization electrodes, and four independent phonon sensors are
    photolithographically patterned. The phonon signal is measured using the Transition
    edge sensor (TES) sensor~\cite{Strauss:2017cuu}.
    Our proposed detector aims to measure both the ionization and phonon energy for
    every event, enabling the simultaneous measurement of both signals.
    This combined information offers an efficient discrimination against the significant
    background of electron recoils originating from the natural $\gamma$ and $\beta$
    radioactivity. Additionally, the ionization signal can be combined
with the phonon signal to provide the true energy of the nuclear recoil and for the
 deduction of quenching factor, Q~\cite{EDELWEISS:2005lsl}. 

\section{Anti-Neutrinos from Nuclear Reactor}
\label{sec:reactor} 
A nuclear reactor is an intense source of electron \antinue s. There are two
primary processes contribute to the generation of \antinue s within the reactor.
 One  of them is the beta decay of fission fragments, predominantly from 
four isotopes $^{235}$U, $^{238}$U, $^{239}$Pu and $^{241}$Pu and the second one is 
the neutron capture process on $^{238}$U. The antineutrinos produced from the beta 
decay of fission fragments have energy up to about 10 MeV, while those produced from 
neutron capture have energy less than 2 MeV. There are on an average six \antinue s 
produced per fission excluding \antinue s produced due to  neutron capture.
 Reactor \antinue s with energy greater than 2.0 MeV are 
easy to detect, whereas measuring low-energy \antinue s poses a significant 
challenge~\cite{Liao:2023kyy}.
In the present work, we have considered the full neutrino flux that includes contribution from fission 
as well as  neutron capture ($^{238}U(n,\gamma) ^{239}U$) process.
 The 
parameterization for \antinue s energy spectra above 2.0 MeV is considered from 
the Huber-Muller model~\cite{Huber:2011wv,Mueller:2011nm}. 
  The low energy \antinue s flux distribution due to
  fission $^{235}$U, $^{238}$U, $^{239}$Pu and $^{241}$Pu is considered from Ref.~\cite{TEXONO:2006xds} and the neutron capture
  process on the $^{238}$U taken from Ref.~\cite{Vogel:1989iv}.  
The relative contribution of each isotope depends on the type of reactor
and its fuel cycle. In this study, reactors with different core
compositions are considered as discussed below.
In the beginning, it is planned to perform measurements with the detector at 4 m from 
the reactor core in the upgraded Apsara (U-Apsara) research reactor facility in Bhabha 
Atomic Research Centre (BARC), India. The U-Apsara reactor has a  compact core with a 
height of about 0.64 m and a radius of about 0.32 m which can operate at a maximum 
thermal power of 3 MW$_{\mathrm {th}}$~\cite{SINGH2013141}. The main advantage of the
U-Apsara reactor is that its core is movable so the measurement can be performed at
different baselines which leads to canceling the reactor as well as detector related systematic
uncertainties. In future, the same 
detector setup can be placed at other reactor facilities such as DHRUVA, 
BARC~\cite{AGARWAL2006747}, Proto-type Fast Breeder Reactor (PFBR), IGCAR, Kalpakkam, 
and VVER, Kudankulam in India~\cite{CHETAL2006852}. The DHRUVA reactor core has 
radius $\sim$1.5 m and height $\sim$3.03 m (defined as an extended 
source)~\cite{AGARWAL2006747}, which can 
operate at a maximum thermal power of 100 MW$_{\mathrm{th}}$ consuming natural 
uranium as fuel. On the other hand, PFBR is relatively a compact source as compared 
to DHRUVA with a dimension of about 1 m  both in radius and height. The PFBR can 
operate at a maximum thermal power of 1250 MW$_{\mathrm{th}}$ with mixed oxide (MOX, 
PuO$_{2}$-UO$_{2}$) as fuel~\cite{CHETAL2006852}. The VVER power reactor has thermal 
power of 3000 MW$_{\mathrm{th}}$ and core has radius $\sim$1.5 m and height 
$\sim$3.03 m (also an extended source). The VVER reactor is a pressurized water 
reactor and uses 3.92$\%$ enriched uranium as a fuel~\cite{AGRAWAL2006812}. 
It can be noted here that the fractional contributions of
each isotope to the reactor thermal power and the parameter
lists used to fit the neutrino energy spectra for the above mentioned reactor
are mentioned in Ref.~\cite{Behera:2020qwf}.
Due to their compact size, U-Apsara and PFBR reactors are the ideal sources to 
utilize the detector set-up for investigating the ASN mixing at short distances. On 
the other hand, at very close distances there are significant background
  contributions from the reactor come into play, impacting the sensitivity of
  sterile neutrino measurements.
The above-mentioned reactors are not only different with respect to 
their sizes and thermal power but also in fuel compositions, as 
detailed in Table~\ref{tab:reactortype}.  
\section{Coherent Neutrino-nucleus Scattering Measurement}\label{sec:cens}
The CE$\nu$NS scattering has been first proposed by 
Freedman~\cite{Freedman:1973yd} within the SM. In this process, the low-energy 
neutrinos scatter off nuclei  which carry only energies up to a few keV. It is very 
hard to measure such low energies ($\sim$ few tens of eV) of recoiling nuclei 
which requires the minimization of uncertainties associated with the relevant 
measurements. The differential CE$\nu$NS scattering cross-section is 
expressed as:
\begin{equation}\label{eq:xsec}
\begin{split}
\frac{d\sigma}{dT}(E_{\nu},T) & = \frac{G_{F}^{2}}{8\pi}
\left[Z(4\mathrm{sin}^2\theta_{W} - 1)+N\right]^2 \\
 & \times A\left(2-\frac{T A}{E_{\nu}^2}\right)|f(q)|^2
\end{split}
\end{equation}  
where $A, N,$ and $Z$ are the mass number, number of neutrons, and number of protons in 
the nucleus, respectively. Further, $E_{\nu}$ is the incident neutrino energy,
$T$ is nuclear recoil energy, with its maximum value denoted as
($T_{\rm  max}(E_{\nu})=2E_{\nu}^{2}/(A+2E_{\nu})$), $G_{F}$ is the 
Fermi coupling constant, $\theta{_W}$ is the weak mixing angle, and $f(q)$ is the 
nuclear form factor for a momentum transfer $q$. For low-energy neutrinos 
($E_{\nu}<$ 50 MeV), the momentum transfer is very small such that $q^2R^2<$1, where 
R is the radius of the nucleus, $f(q) \sim 1$. At small momentum transfers, the 
scattering amplitude from individual nucleons is in phase and added coherently, 
which leads to the increase of cross-section.The weak mixing angle
$\mathrm{sin}^{2}\theta_{W}$ has been experimentally determined to be 0.23867 $\pm$ 
0.00016 $\sim$1/4~\cite{Erler:2004in}. Hence, the cross-section is proportional to 
$N^2$. Although the scattering cross-section is enhanced by the number of nucleons, 
it depends on the measurement of very low energies of recoiling nuclei. The recoil 
energy of the nucleus depends on its mass, which decreases with the increase in the 
mass of target nuclei. For instance, with a neutrino energy of 1 MeV, the 
maximum recoil energy is about 20 eV and 50 eV for Ge and Si targets, respectively. 

\begin{figure}[t]
\centering
\includegraphics[width=0.5\textwidth]{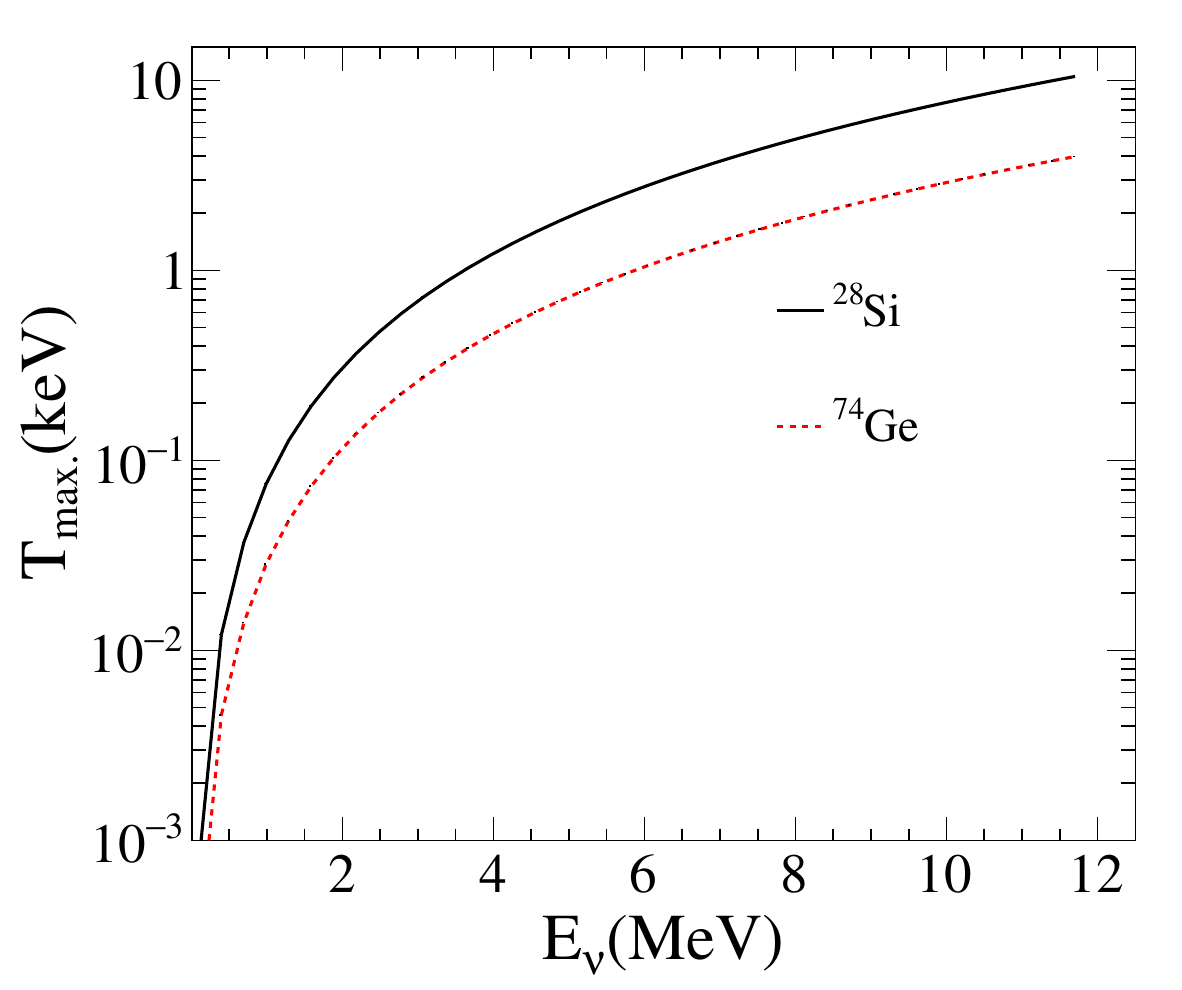}
\caption{ \label{fig:recoilEnergy} Maximal recoil energy of Si and Ge nuclei 
as a function of energy of the reactor \antinue s.} 
\end{figure}

\begin{table*}
 \begin{center}
\caption{\label{tab:calEvt}{Expected events with different detectors placed at 4m 
from U-Apsara reactor}}
\begin{tabular}{ ccc}
Threshold (eV) & Si detector & Ge detector \\
  \hline
 20.0 & 1443.0 & 2017.0  \\
 50.0 & 1256.0 & 1501.0  \\ 
 100.0 & 1038.0 & 982.0 \\
 \hline
\end{tabular}
\end{center}
\end{table*}
The maximal recoil energy of different target nuclei (Si, Ge) considered in
this study is shown in Fig.~\ref{fig:recoilEnergy} as a 
function of reactor \antinue energy. It can be observed from 
Eq.~\ref{eq:xsec} that the cross-section is maximum at zero recoil energy and it 
decreases with the increase of $T$. Hence, a detector with a higher threshold 
energy for detecting the signal leads to a lesser number of events. Therefore, it is
very tough to select the type of detector for the measurement of 
such a cross-section. Because of the low energy of the antineutrinos, the 
recoil energy deposited in the detector is up to a few keV.
\section{Expected Event Rate in Detector}
\label{sec:expect}
\begin{figure}[h]
\centering
\includegraphics[width=0.5\textwidth]{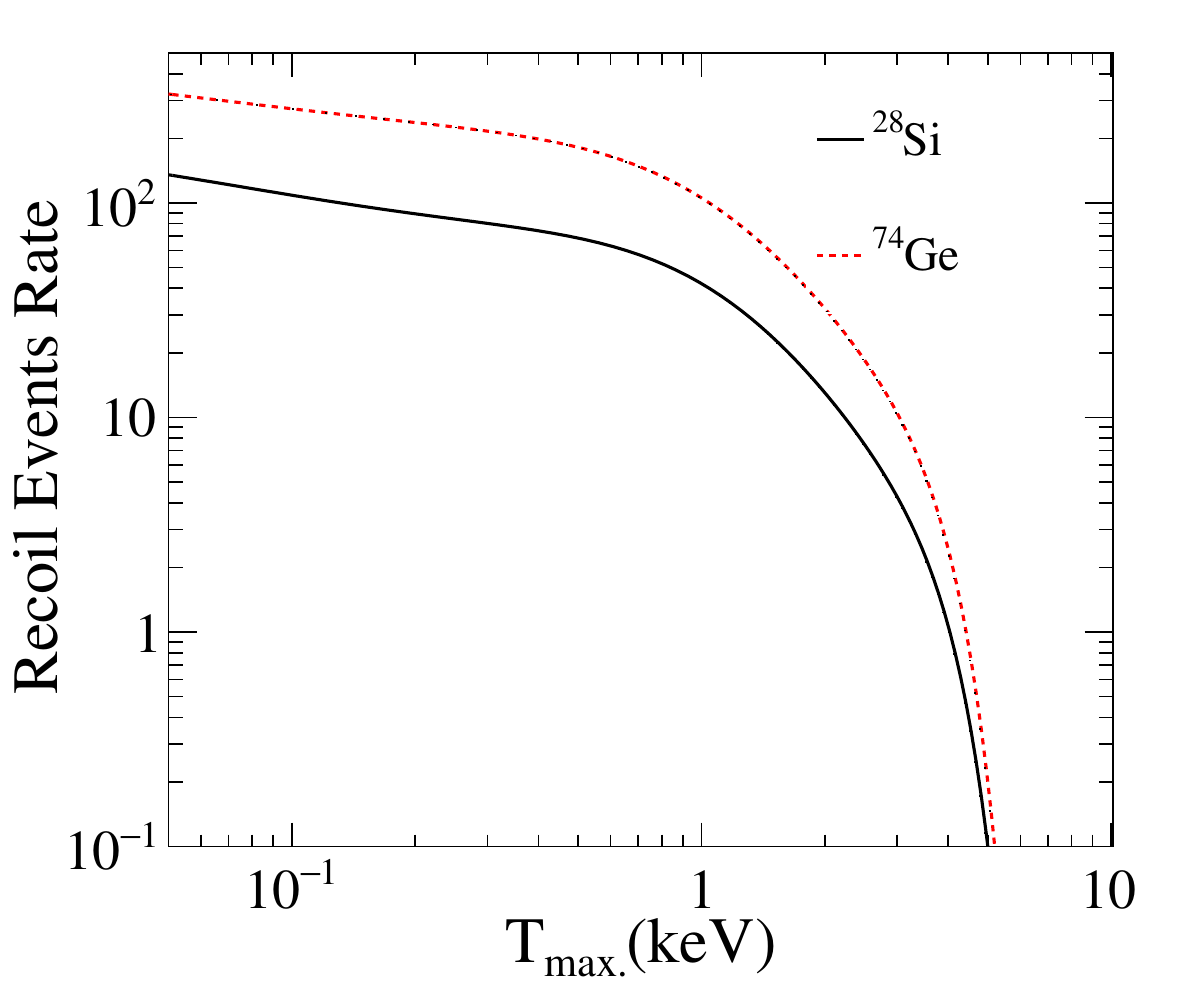}
\caption{ \label{fig:diffrate} Differential event rate variation with maximal recoil energy 
using Si and Ge detectors.} 
\end{figure}

The CE$\nu$NS reaction cross section per unit detector mass could be up to two orders of 
magnitude greater than that for IBD process, potentially allowing for detectors in the kilogram 
range. The expected signal event rate due to CE$\nu$NS is given by
\begin{equation}\label{evt}  
\begin{split}
\frac{dN}{dT} &=t\lambda_{0} \frac{M_{\text{det}}}
 {A}\int_{E_{\nu}^{\rm min}}^{E_{\nu} ^{\rm max}}\lambda(E_{\nu})dE_{\nu}\\
&\int_{T_{\rm min}}^{T_{\rm max}}\left(\frac{d\sigma}
 {dT(E_{\nu})}\right) dT(E_{\nu}) ,
\end{split}
\end{equation}
\noindent where $M_{\text{det}}$ is the mass of the detector, $t$ is the 
time duration of data taking, $\lambda_{0}$ is the total \antinue flux and, 
$\lambda(E_{\nu})$ is the \antinue energy spectrum. For a given experimental setup, 
the detector threshold decides the minimum recoil energy of the nuclei.
Due to the lower abundance of \antinue s with energies exceeding 6.0 MeV,
  we have considered \antinue s with a maximum energy of approximately 6.0 MeV for
  the analysis.
It can be noted here that the expected number of events in the detector is 
estimated by considering the contributions from each stable isotope of the element 
of the detector (Si or Ge)
weighted by its natural abundance. Figure~\ref{fig:diffrate} shows the differential 
event rate dependence of maximal recoil energy for Ge and Si detectors. For a given recoil 
energy, the number of events increases with the target mass number due 
to an increase in the number of neutrons. However, the number of events decreases with the 
increase of recoil energy. Consequently, it becomes imperative to either augment
  the target mass or the energy of neutrinos for targets with lower mass numbers.
As previously mentioned, the
expected number of events also depends on the detector threshold. 
The number of events expected in the detector of mass 10 kg is mentioned in 
Table~\ref{tab:calEvt}. Events are estimated considering the detector is placed at 4m 
distance from the U-Apsara reactor core for an exposure of 1 year.
\section{Neutrino Oscillation Probability in Short Baseline Experiment}%
\label{sec:osciprob}
The CE$\nu$NS process offers a distinct advantage in assessing ASN
  mixing sensitivity compared to the IBD process, owing to its higher cross-section.
  This allows for the use of smaller size detectors in CE$\nu$NS measurements,
  thereby minimizing the uncertainty in the neutrino's path length.
  In the Standard Model, three active neutrino flavors
  ($\nu_{e}$, $\nu_{\mu}$, $\nu_{\tau}$) are present, and their conversion to
  mass eigenstates is described by the Pontecorvo-Maki-Nakagawa-Sakata (PMNS)
  unitary matrix)~\cite{Maki:1962mu}. Numerous global efforts are underway to
  study neutrino oscillation phenomena and measure oscillation parameters for the
  three-generation model~\cite{Super-Kamiokande:2010orq,SNO:2011hxd,KamLAND:2008dgz,
DayaBay:2012fng,RENO:2012mkc,DoubleChooz:2012gmf,K2K:2006yov,MINOS:2012wcj,
T2K:2011qtm}. Beyond these, experiments are exploring to either find out
  or exclude the existence of sterile neutrinos, which has no analogous
  SM gauge interactions. However, its presence can affect the standard neutrino 
oscillations.
  
 Firstly, a standard neutrino could oscillate into an 
undetectable sterile neutrino, leading to a reduction of the observed event rate 
within the detector. Secondly, the mass eigen-state ($\nu_{4}$ with mass 
$\mathrm{m_{4}}$) primarily associated with the sterile neutrino would enhance the 
transformation probability between standard neutrinos, leading to the detection of a 
neutrino flavor that is not emitted by the source. The experiments looking for a 
reduction of the interaction rate are called ``disappearance" experiments while the 
ones seeking an enhanced neutrino conversion are called ``appearance" 
experiments.

  In the context of ASN oscillation, the PMNS matrix extends to 3$+$1 from
  the standard 3 generations, with "3" representing active neutrinos and
  "1" representing a sterile neutrino ($\nu_{s}$). The specifics of the
  rotation order and mixing matrix elements can be found in
  Ref.~\cite{Behera:2019hfs}. The 3$+$1 generation oscillation model reduced to
  a two-flavor framework for small mixing angles ($\theta_{14}$) and short
  source-to-detector distances ($<$ 100 m).
Then the $\antinue$ survival probability can be approximated as  
\begin{equation}
\label{eq:prob}
P_{\antinue\antinue}(E_{\nu},L) \simeq 1-\sin^22\theta_{14}\sin^2\left(\frac{1.27 
\Delta m^2_{41}{L}}{E_{\nu}}\right),
\end{equation}
where $E_{\nu}$ is the \antinue energy (in MeV), $L$ is the path length (in m) 
between the source and the detector, and $\Delta m^2_{41}$ is the squared masses 
difference (in eV$^2$) between the two (anti)neutrino mass eigen-states. 
The ASN oscillation parameters $\Delta m^2_{41}$ and 
$\sin^22\theta_{14}$ are represented by
\begin{equation}
\Delta m^2_{41}  =  m^{2}_{4}- m^{2}_{1}~;~~
\sin^22\theta_{14}  =  4 |U_{e4}|^2(1 - |U_{e4}|^2), 
\end{equation}
where $U_{e4}$ = $\sin\theta_{14}$, one of the elements of unitary mixing matrix.
 The combined analysis of data obtained by NEOS and DANSS 
collaborations provides the present best-fit values of ASN
oscillation parameters as $\Delta m^2_{41} \simeq $ 1.30 eV$^{2}$ and 
$\sin^22\theta_{14} \simeq$ 0.049 ~\cite{Gariazzo:2018mwd}. These findings
  align with the results of a global analysis~\cite{Dentler:2018sju}. 
With these values of ASN mixing parameters, experimentally, the possible existence of 
sterile neutrinos at short baseline can be observed by finding the distortions of the 
$\antinue$ energy spectrum which is otherwise absent in three active neutrino 
oscillation. 

\section{Simulation Procedure}
\label{sec:simul}              %
In the present study, the potential of different detectors has been explored for 
finding the ASN oscillation sensitivity by using the \antinue s produced from 
various types of reactor facilities as detailed in Table~\ref{tab:reactortype}.
It is important to note that the energy spectrum of the \antinue s produced
  from these reactors varies across isotopes. Consequently, the number of
  \antinue s isn't solely dependent on thermal power but also on fuel compositions. 

The spatial 
variation of \antinue s  flux due to the finite-size cylindrical shape reactor core is considered 
 which can be parameterized as~\cite{Glasstone_Sesonske}, 
\begin{equation}
\phi = \phi_{0}~J_{0}(2.405 r/R)~cos(\pi z/H)
\end{equation}   
where $\phi_0$ represents the flux at the center of the reactor core taken as the 
vertex position, $R$, and $H$ are the physical radius and height of the cylindrical 
reactor core, respectively, $J_0$ is the zeroth-order Bessel function of the first kind 
with $r$ ($0\leq r\leq R$) and z ($ 0 \leq z \leq H$). In this analysis, events 
are estimated using the reactor antineutrinos flux and CE$\nu$NS cross-section, as 
previously outlined. The detector response to the recoil energy  spectrum is 
incorporated by assuming a standard Gaussian form with the standard 
deviation($\sigma$) for the energy resolution as given by
\begin{equation}
R^{\text {res}}(T_{m},T) = 
\frac{1}{{\sqrt{2\pi}}\sigma} \exp(-\frac{(T_m - T)^2}{2\sigma^2})\,.
\label{Ereso}
\end{equation}
where $T$ and $T_m$ are the simulated true and observed recoil energy of nuclei, 
respectively. The detector resolution is considered as $\sigma/T ~\sim$ 10$\%$/$\sqrt{T}$. The 
 recoil energy spectrum generated due to \antinue s induced events is distributed with 
variable bin widths such that the minimum number of events in each bin is $\geq$ 5. The 
number of events in $i$-th energy bin after folding the detector resolution is given 
as
\begin{equation}
N_{i}^{r} = \sum_{k} K_{i}^{k}(T^{k}) n_{k}
\end{equation}
The index $i$ corresponds to the measured energy bin and $N_{i}^{r}$ represents 
the number of reconstructed events, $k$ is summed over the true recoil energies of 
nuclei and $n_{k}$ is the number of events in $k$-th true energy bin. Further, 
$K_{i}^{k}$ is the integral of the detector resolution function over the $T$ bins 
 which is given as
\begin{figure*}[t]
\includegraphics[width=0.46\linewidth]{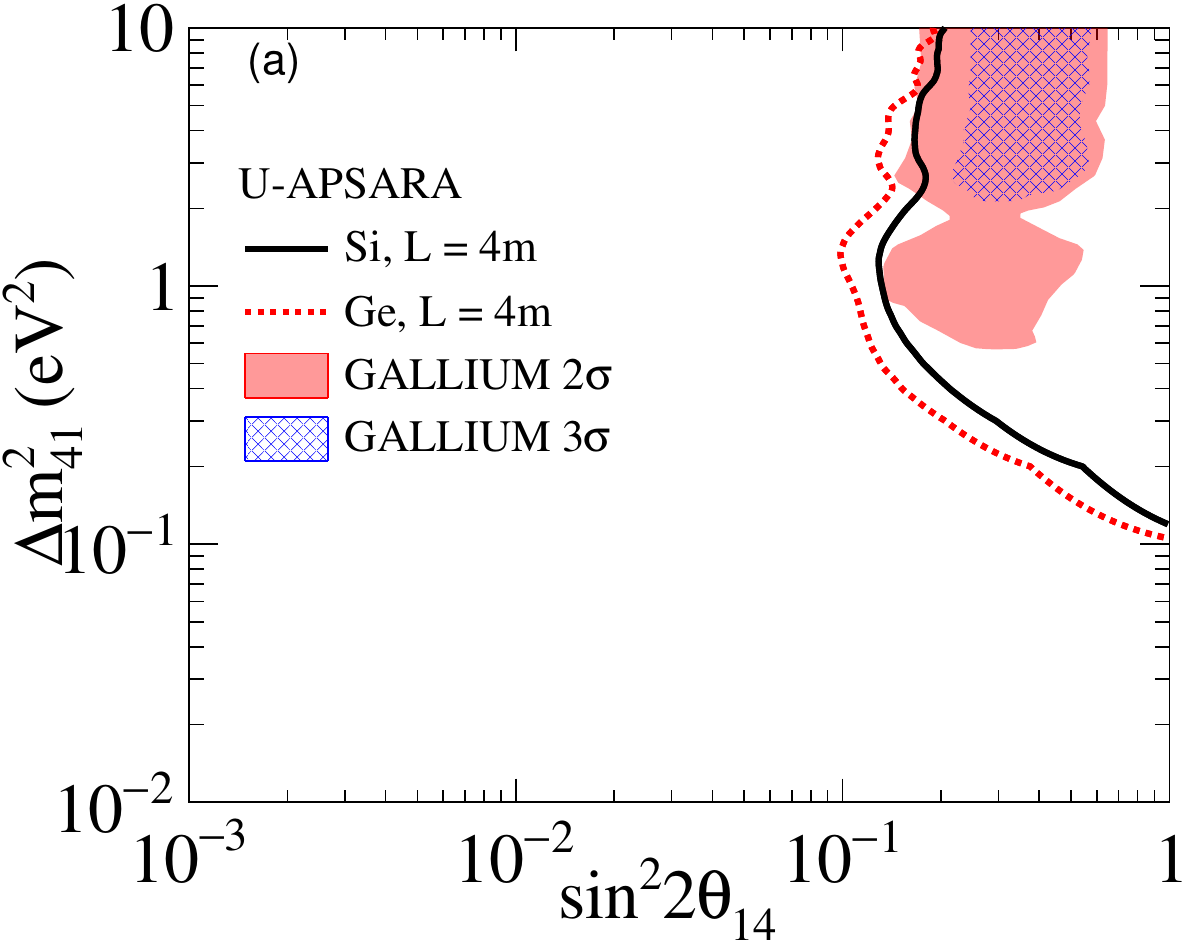}
\includegraphics[width=0.45\linewidth]{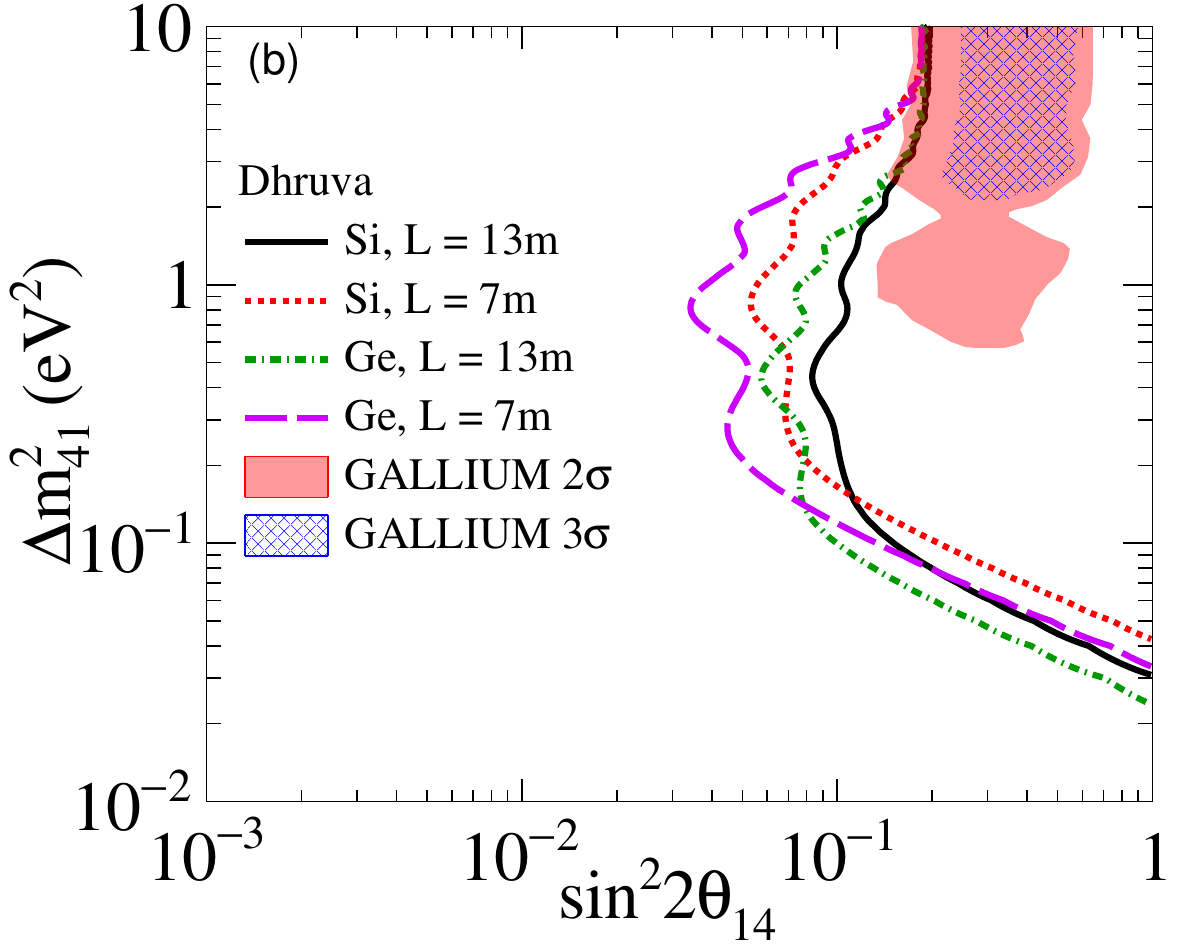}
\includegraphics[width=0.45\linewidth]{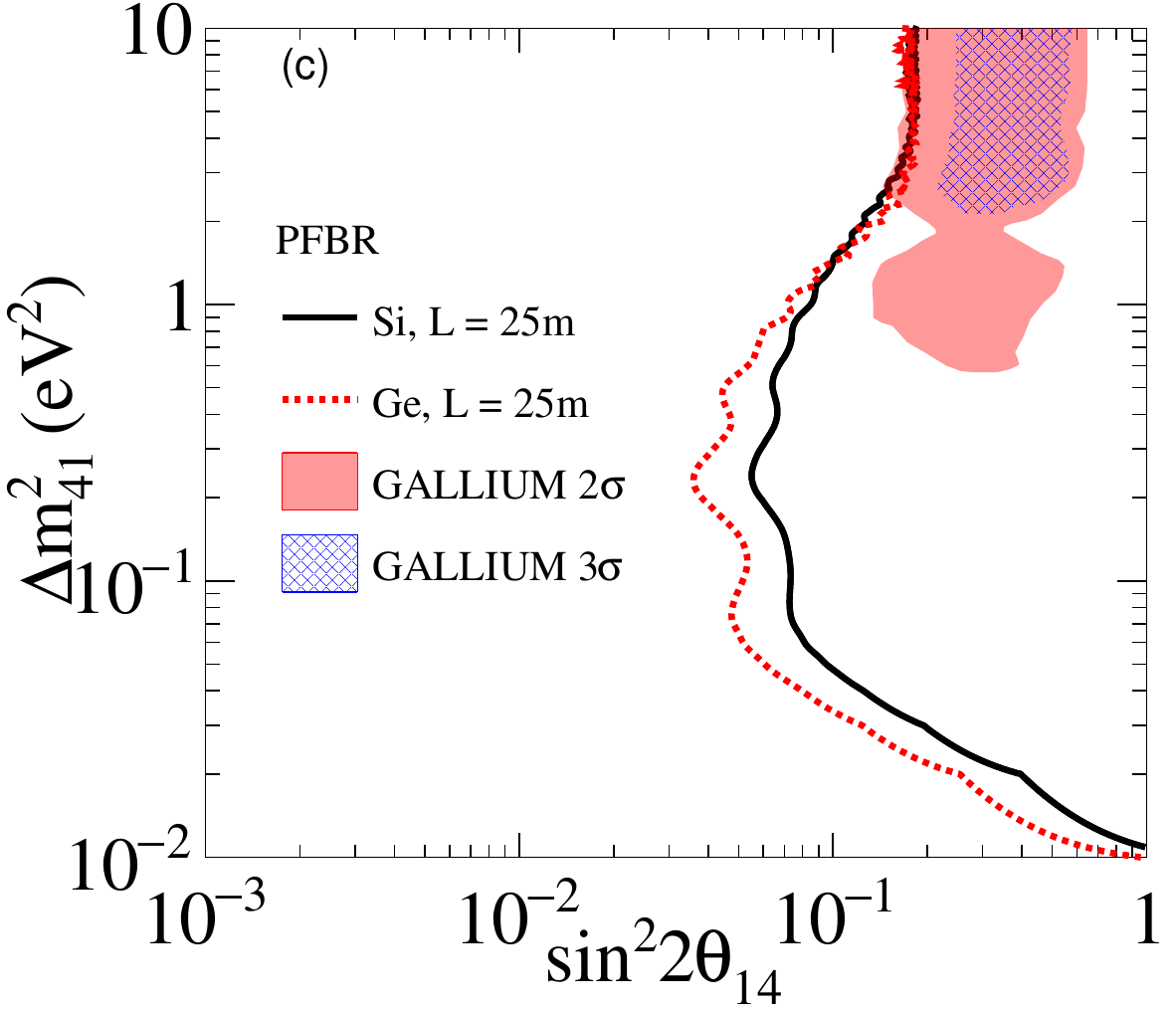}
\hspace{1.2cm}
\includegraphics[width=0.45\linewidth]{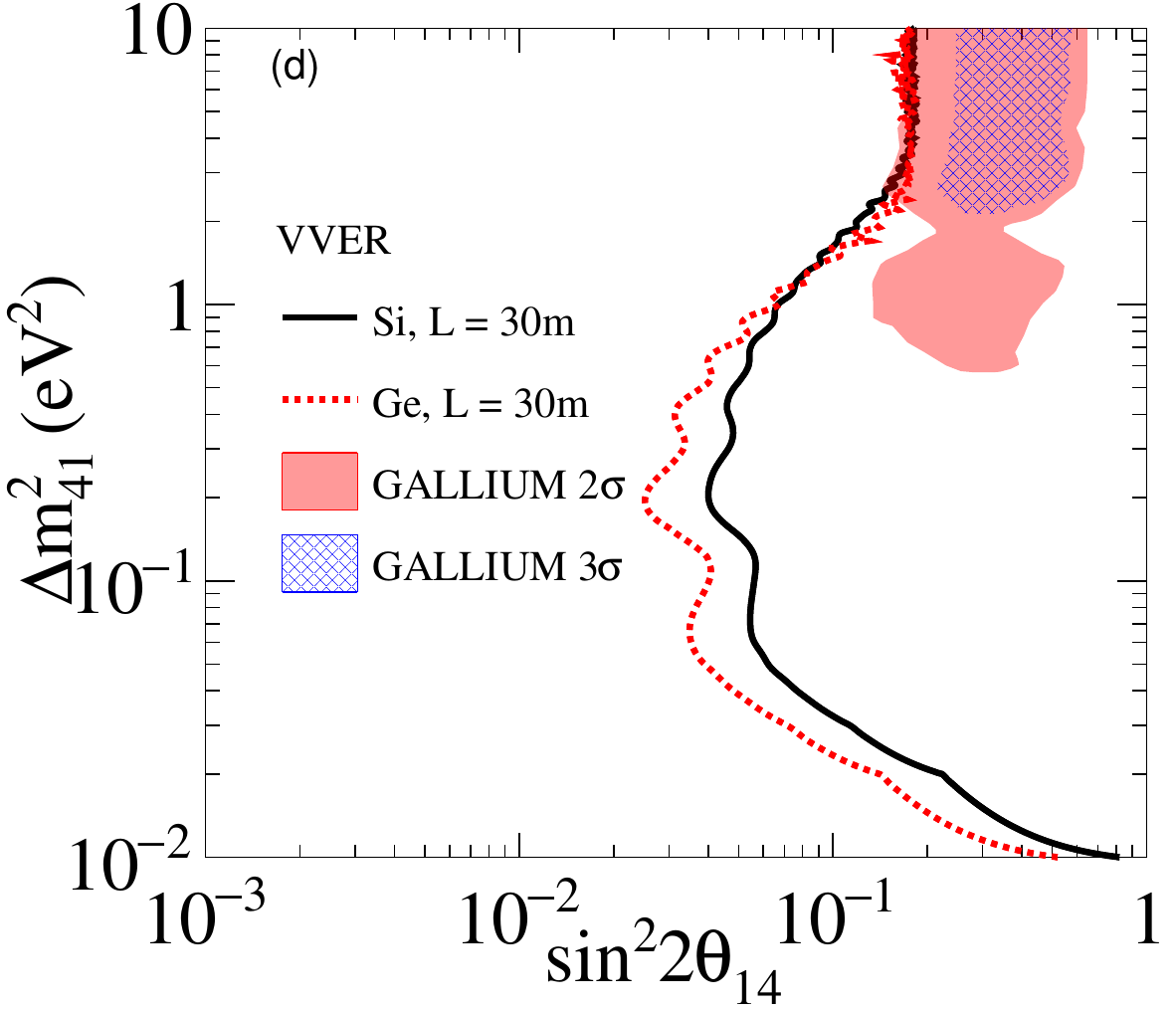}
\caption{ \label{fig:diffreactors} 
The comparison of ASN mixing sensitivity at 95 $\%$ C.L. for the Ge and Si 
detectors placed at fixed distances of 4 m, 13 m, 25 m, and 30 m from the (a) U-Apsara, 
(b) DHRUVA, (c) PFBR and (d) VVER reactors, respectively.} 
\end{figure*}

\begin{equation}
K_{i}^{k} = \int_{T_{L_{i}}}^{T_{H_{i}}} dT \frac{1}{\sqrt{2\pi\sigma_{T}^{2}}} 
e^{-{\frac{\left(T^{k}-T_{m}\right)^{2}}{2\sigma_{T}^{2}}}}
\end{equation}
The integration is performed between the lower and upper boundaries of the measured 
energy ($T_{L_i}$ and $T_{H_i}$) bins. In the present analysis, we have assumed 
80$\%$ for the detection efficiency, 90$\%$ as the fiducial volume of the detector, 
and 70$\%$ reactor duty cycle for a total exposure of 1 year. The production 
points of \antinue s inside the reactor core are generated randomly using a 
Monte-Carlo method whereas a point detector is considered due to its compact size.  
\section{Extraction of Active-Sterile Neutrino Mixing Sensitivity}
\label{sec:chisq}  
The detector sensitivity to the ASN mixing can be extracted by knowing the neutrino 
energy spectrum, flux, and its cross-section accurately. A total number of 
neutrino-induced events expected within the detector can be estimated using the 
procedure mentioned above for a given oscillation hypothesis and it can be compared 
with the actual measured events. For this purpose, a statistical analysis between the 
predicted and expected event distribution by simulation is carried out in order 
to quantify the sensitivity of the detector to the ASN mixing parameters 
$\theta_{14}$ and $\Delta m^2_{41}$ for a given exposure. The detector response is 
folded in both simulation predicted as well as the expected events. The sensitivity 
to the sterile neutrino mixing parameters is extracted by 
estimating the $\chi^2$ method.
The exclusion limit is obtained for each value of $\Delta m_{41}^2$ by scanning over 
the various values of $\sin^2 2\theta_{14}$ and identifying the boundary of the 
corresponding $\chi^2$ (e.g. $\chi^2$ = 5.99 for 95.0\% confidence limit (C.L.)). The 
definition of $\chi^{2}$ is taken from Ref.~\cite{Gonzalez-Garcia:2004pka} and 
given as
\begin{equation}
\chi^{2} =\sum_{n=0}^{N} \bigg(\frac{ 
N_{n}^{th}-N_{n}^{ex}}{\sigma(N_{n}^{ex})}\bigg)^{2} + \sum_{i=0}^{k} \xi_{i}^{2},
\label{eq:chi1}
\end{equation}%
where, $n$ is the number of energy bins with variable widths,
$N_{n}^{ex}$, $N_{n}^{th}$ are the number 
of events obtained from the simulations with oscillation (expected) and without 
oscillation (theoretically predicted) events, respectively. The theoretically 
predicted events $N_{n}^{th}$ are calculated considering the reactor \antinue~ flux, 
the  CE$\nu$NS cross-section, detection efficiency, and energy resolutions of the 
detector as mentioned earlier. The $N_{n}^{ex}$ is estimated by folding the
oscillation probability on 
$N_{n}^{th}$ along with the detector resolution. The $N_{n}^{th}$ carries the 
information about the systematic uncertainties given by
\begin{equation}
N_{n}^{th}=N_{n}^{'th}\bigg(1+\sum_{i=0}^{k}\pi_{n}^{i}\xi_{i}
\bigg)+\mathcal{O}(\xi^{2})
\label{eq:chi2}
\end{equation}
with $\pi_{n}^{i}$ being the strength of the coupling between the pull variable 
$\xi_{i}$ and $N_{n}^{'th}$. The $\chi^2$ is minimized with respect to pull 
variables $\xi_i$ and it is estimated by considering four sources of systematic 
uncertainties. It includes  3.0$\%$ normalization uncertainty which arises due to 
reactor total neutrino flux, number of target atoms, and detector efficiency, 
uncertainty due to nonlinear energy response of the detector taken as 1.0$\%$, and, 
uncertainty in the energy calibration given as 0.5$\%$. In addition, the 
uncorrelated experimental bin-to-bin systematic error of 2.0$\%$ is also considered 
which could result due to the insufficient knowledge of other sources of background.
\section{Active-Sterile Neutrino Mixing Sensitivity}
\label{sec:results}
Earlier, the sensitivity of ASN mixing was explored using the
  ISMRAN detector setup, assuming a fixed distance between the detector
  and the reactor core and \antinue s will be detected through the IBD 
process~\cite{Behera:2019hfs,Behera:2020qwf}. The present analysis has been carried 
out by varying both the reactor and detector-related parameters with the neutrinos 
detected through CE$\nu$NS process.  The comparison of active to sterile
  neutrino oscillation parameters' sensitivity of various types of detectors 
 is performed by employing $\antinue$s produced from different 
types of reactors as mentioned in Table~\ref{tab:reactortype}.
\subsection{With Different Types of Detector}
In contrast to the IBD scenario, there is a wider range of detector materials available 
in which CE$\nu$NS can be measured. While the CE$\nu$NS scattering cross section 
increases with the number of nucleons present in the target material, the recoil
  energy of the nuclei decreases with mass.
\begin{figure*}[t]
\includegraphics[width=0.45\textwidth]{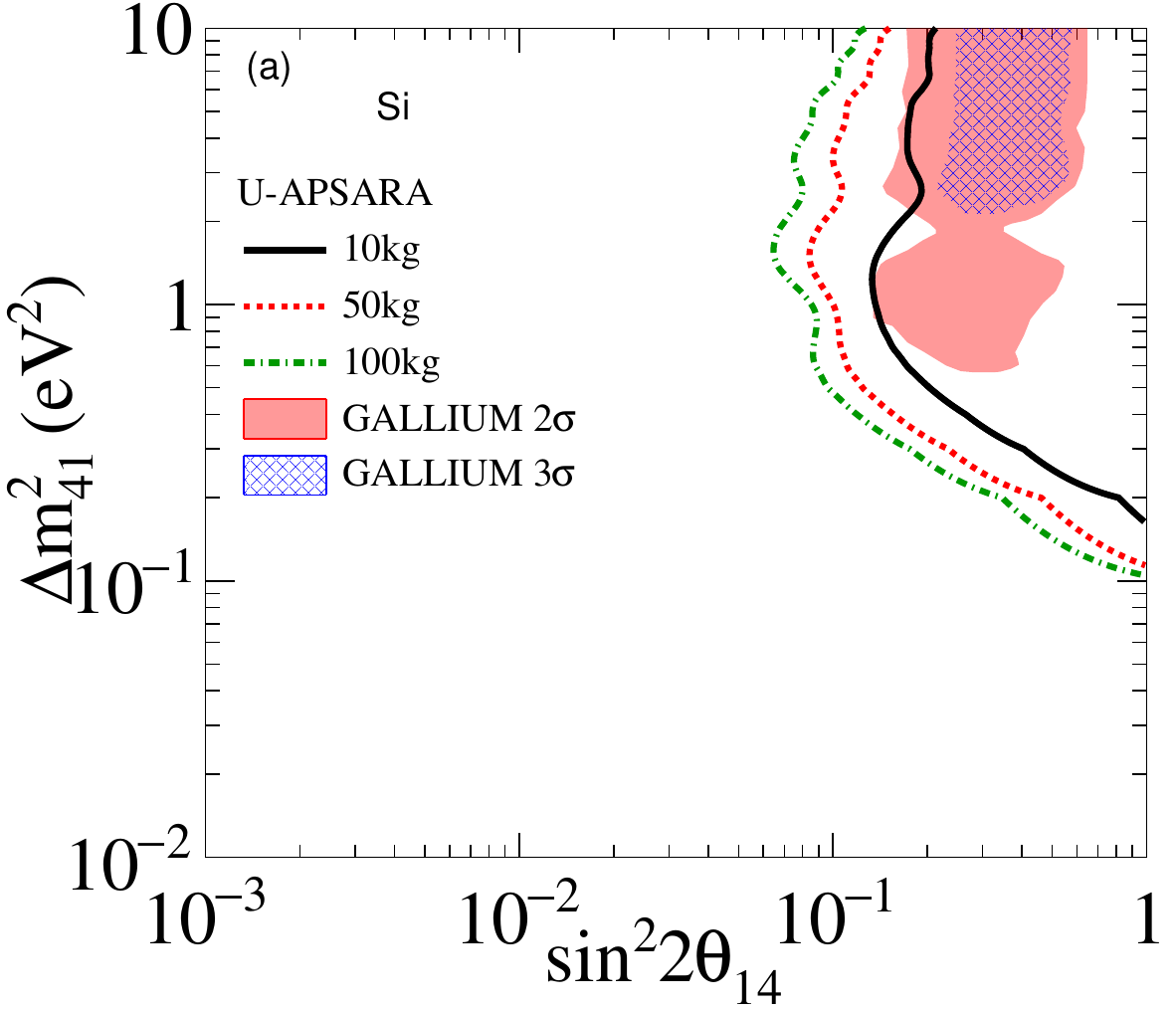}
\includegraphics[width=0.45\textwidth]{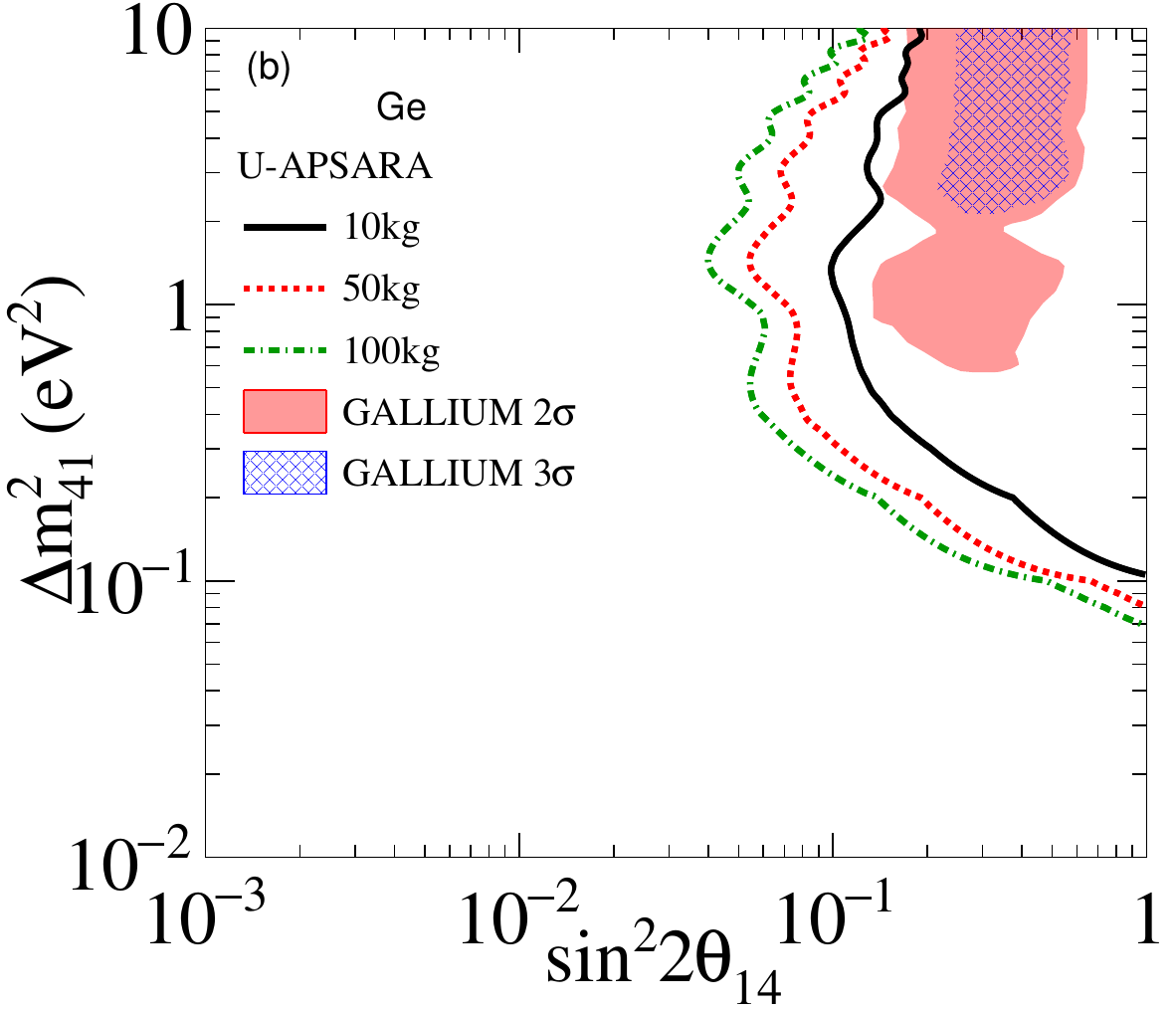}
\caption{ \label{fig:SenApsaraDetmass} The comparison of ASN mixing sensitivity at
  95 $\%$ C.L. for different detector mass of the (a) Si and (b) Ge detectors placed
  at distance of 4 m from the U-Apsara reactor}. 
\end{figure*}
\begin{figure*}[t]
\includegraphics[height=0.4\textwidth]{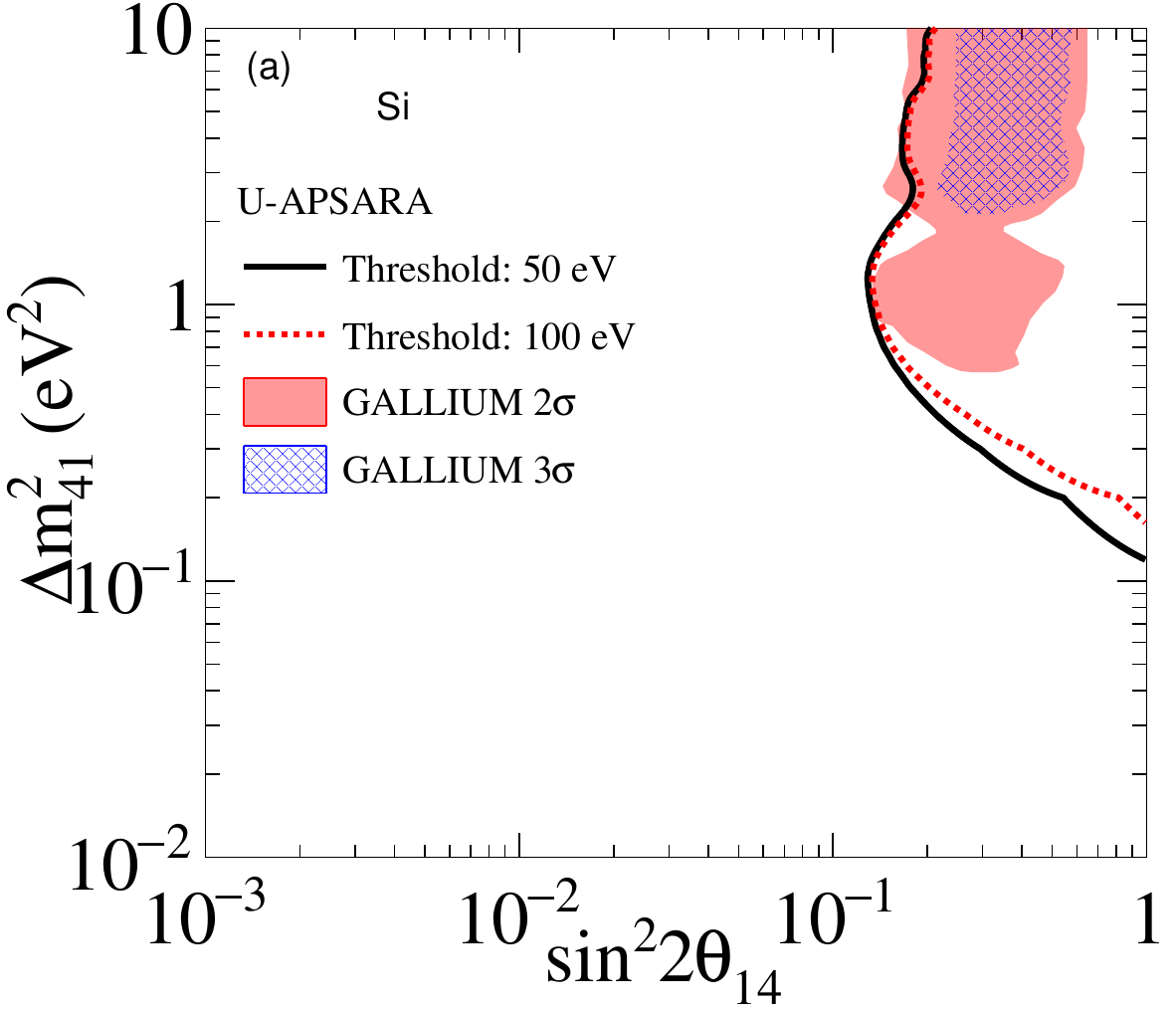}
\includegraphics[height=0.4\textwidth]{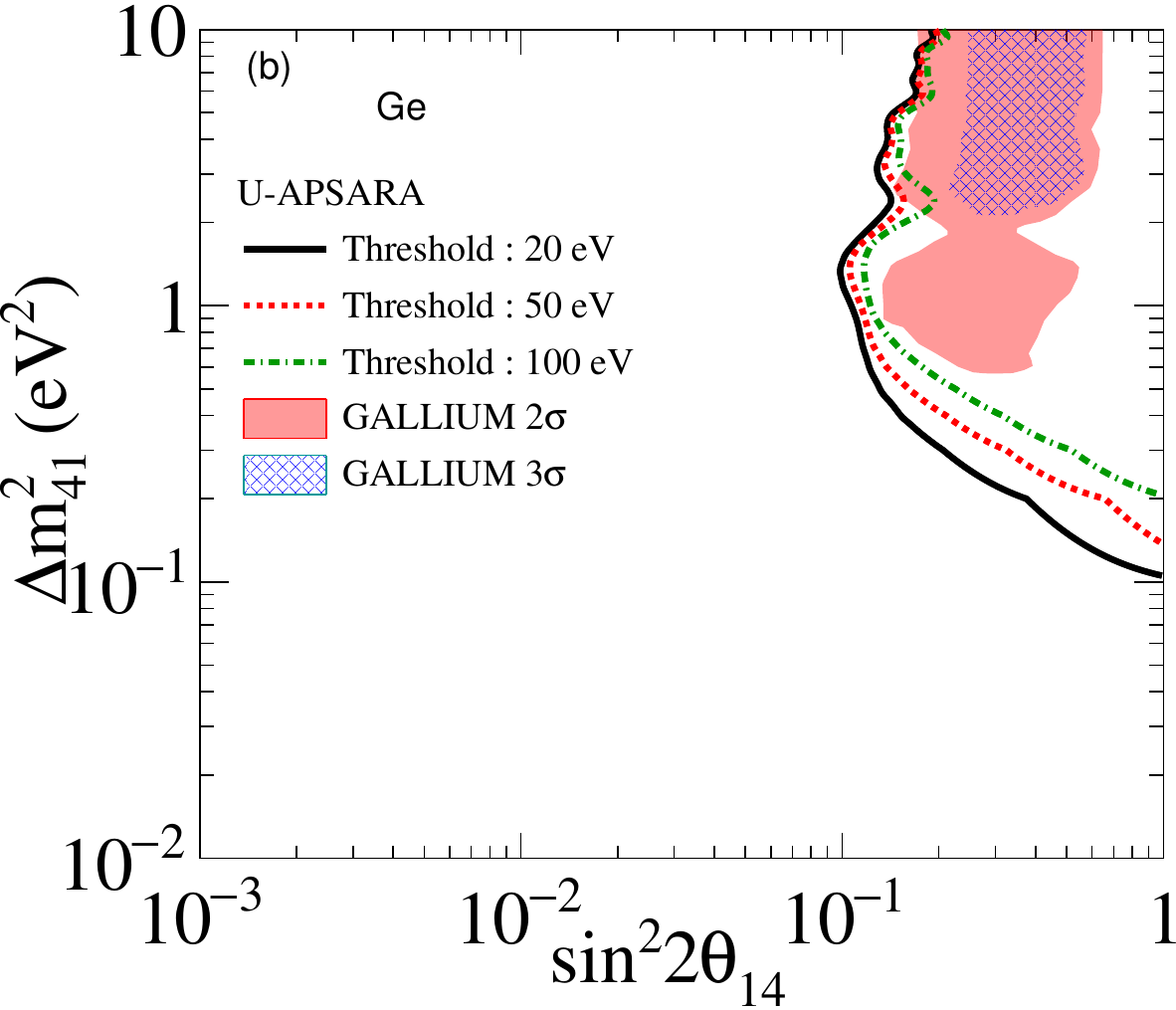}
\caption{ \label{fig:SenpDhruvaTh} The comparison of ASN 
mixing sensitivity at 95 $\%$ C.L. for different detector threshold for the (a) Si and (b) 
Ge  detectors placed at 4 m from the U-Apsara reactor facility.} 
\end{figure*}

The ASN mixing 
sensitivity of the detector has been studied considering commonly employed detector 
materials, such as Germanium (Ge) and Silicon (Si), each having a payload of 10 
kg at a given reactor thermal power and reactor core to detector distance. Figure 
~\ref{fig:diffreactors} shows the extracted results in sin$^22\theta_{14}$ -- 
$\Delta m^2_{41}$ plane at 95.0 $\%$ C.L for an exposure of 1.0 year. Results are 
presented for different distances between the detector and the core of the reactor 
based on the accessibility conditions of different reactors. We assume a recoil energy 
threshold of 20.0 eV and 50.0 eV for Ge and Si detectors, respectively. Allowed 
regions for the Gallium anomaly are also depicted in the same 
figure~\cite{Kopp:2013vaa}. Figure ~\ref{fig:diffreactors}(a) shows the sensitivity 
by placing the detectors at 4 m from the U-Apsara research reactor core. Figure 
~\ref{fig:diffreactors}(b) shows the sensitivity of the detector by positioning it at 
7.0 m and 13.0 m from the DHRUVA reactor core. Figure~\ref{fig:diffreactors}(c) shows the 
sensitivity of the detector at the PFBR reactor which is at a distance of 25.0 m from the 
detector, and Figure ~\ref{fig:diffreactors}(d) shows the sensitivity for the 
detectors placed at 30 m from the VVER power reactor core. The shape of the 
sensitivity curve in the region of low values  $\Delta m^2_{41}$ ( $\lesssim {1.0}{~eV^2}$)
 shows a linear dependence between sin$^2 2\theta_{14}$ and 
$\Delta m^2_{41}$ in the logarithmic scale. This may happen as typical neutrino 
oscillation lengths are much larger as compared to the size of the detector, and the 
$\antinue$ survival probability mentioned in Eq.~\ref{eq:prob} is approximately given 
by $P_{\nu_e\nu_e}(E_{\nu},L) \approx 1 - C\sin^2 2\theta_{14} \times \left(\Delta 
m^2_{41}\right)^2$, where $C$ is a constant. In the region with higher $\Delta 
m^2_{41}$ values, the systematic uncertainties related to the neutrino source 
dominate over the statistical uncertainties. The detector energy resolution flattens 
the high frequency oscillation-induced deformations significantly, resulting in the 
gradual decrease of the shape-discriminating power. This leads to the event
  distribution, with and without oscillation, overlaps except for a constant
  normalization factor. It is also observed that, by placing the detector at a 
distance in the range of 15--30 m in the case of PFBR and VVER reactors, leads to a 
greater parts of the region can be excluded in 
the sin$^22\theta_{14}$ -- $\Delta m^2_{41}$ plane. Also, the maximum sensitivity of 
the detector shifts to a lower value of $\Delta m^2_{41}$ ($\sim$~0.03 $eV^2$) due to 
the increase in source-to-detector distance. Notably, the Ge detector demonstrates
  superior ASN mixing sensitivity compared to the Si detector when $\Delta m^2_{41}$
  is less than or equal to 1.0 $eV^2$. However, for higher values of
$\Delta m^2_{41} > 1.0~ eV^2$, both the 
detectors exhibit nearly identical sensitivity to the mixing angle  sin$^22\theta_{14}$. 
Both Si and Ge detectors exclude the allowed region of the GALLIUM experiment at C.L. 
of 2$\sigma$. The detector sensitivity also depends on uncertainty in neutrino path length 
which occurs due to spatial variation of \antinue s flux and
finite size of the detector.
By smearing the \antinue s both vertex position in the reactor core (U-Apsara) and the detection
position, it is observed that there is a maximum variation of about 2$\%$ in the mixing angle 
 sin$^22\theta_{14}$ at  $\Delta m^2_{41} > 5.0~ eV^2$ as compared to fixed path length. However, at
lower $\Delta m^2_{41}(\leq 1.0~eV^2)$, there is a minimal effect to the ASN oscillation parameters.
\subsection{With Different Masses of the Detector}
The expected number of events decreases with a decrease in the target mass number. 
Therefore, to enhance the sensitivity of the detector, it is required to increase the 
neutrino flux, exposure time, or the target mass. At a given reactor power, in order 
to increase the CE$\nu$NS interaction rate the actual target mass number of the 
detector is more crucial. Consequently, increasing the detector mass is a 
prudent choice, even if it adds some complexity to handling. Taking into account 
the backgrounds, it is observed that detectors in CE$\nu$NS experiment exhibit a mass advantage of 
one order of magnitude, assuming that measurements of eV-scale recoil thresholds are 
feasible and a signal-to-background ratio of about 1 can be achieved in the 
experiment~\cite{Bowen:2020unj}. Figure~\ref{fig:SenApsaraDetmass} shows the ASN 
mixing sensitivity in sin$^22\theta_{14}$ -- $\Delta m^2_{41}$ plane at 95$\%$ C.L. 
for an exposure of 1 yr for different mass of the detector. 
Figure~\ref{fig:SenApsaraDetmass}(a) shows for Si detector and 
Fig.~\ref{fig:SenApsaraDetmass}(b) for Ge detector, both with masses 10 kg, 50 kg, 
and 100 kg. It can be noted here that, sensitivities are extracted considering a 
threshold of 100 eV and 20 eV for silicon and germanium detector, respectively. At a 
given detector mass at higher $\Delta m^2_{41}$ the sensitivity in 
sin$^22\theta_{14}$ is more as compared to lower $\Delta m^2_{41}$. The sensitivity 
improves in all regions of sin$^22\theta_{14}$ with an increase in detector mass as 
the number of expected events increases. It is evident that the Ge detector excludes 
both allowed regions of the GALLIUM experiment.
\subsection{With Various Detector Thresholds}
As previously discussed, the recoil energy of nuclei decreases
  as the mass number of the detector increases.
Therefore, detectors with low thresholds, such as Ge and Si can perform 
nuclear recoil discrimination down to the eV scale energy threshold. Such detectors will 
be more useful if we can observe the recoil energies as low as a few tens of eV. In
  the context of CE$\nu$NS cross-section, as given in Eq.~\ref{eq:xsec}, the
nuclear recoil energy $T$ serves as the relevant observable. For a
given neutrino energy, there is a limit to the 
maximum recoil energy. The cross-section is maximum for lower values of $T$, 
therefore, the total number of observed event rate sensitivity dependent on the 
low-energy threshold for nuclear recoil $T_{min}$. A comprehensive study
has been performed to assess the impact of detector threshold on
ASN mixing sensitivity, considering both Si and Ge detectors with a
mass of 10 kg placed at a distance of 4 m from
U-Apsara reactor core. Figure ~\ref{fig:SenpDhruvaTh} shows the sensitivity
in sin$^22\theta_{14}$ --$\Delta m^2_{41}$ plane at 95 $\%$ C.L for
detector with  different threshold energies. The 
sensitivity of the detector at lower $\Delta m^2_{41}$ is better at the lower detector 
threshold. However, for $\Delta m^2_{41} \geq 1.0~ eV^2$, the detector threshold has 
minimal impact on ASN mixing sensitivity.
\subsection{With Different Resolution of Detector}
\begin{figure}
\centering
\includegraphics[width=0.45\textwidth]{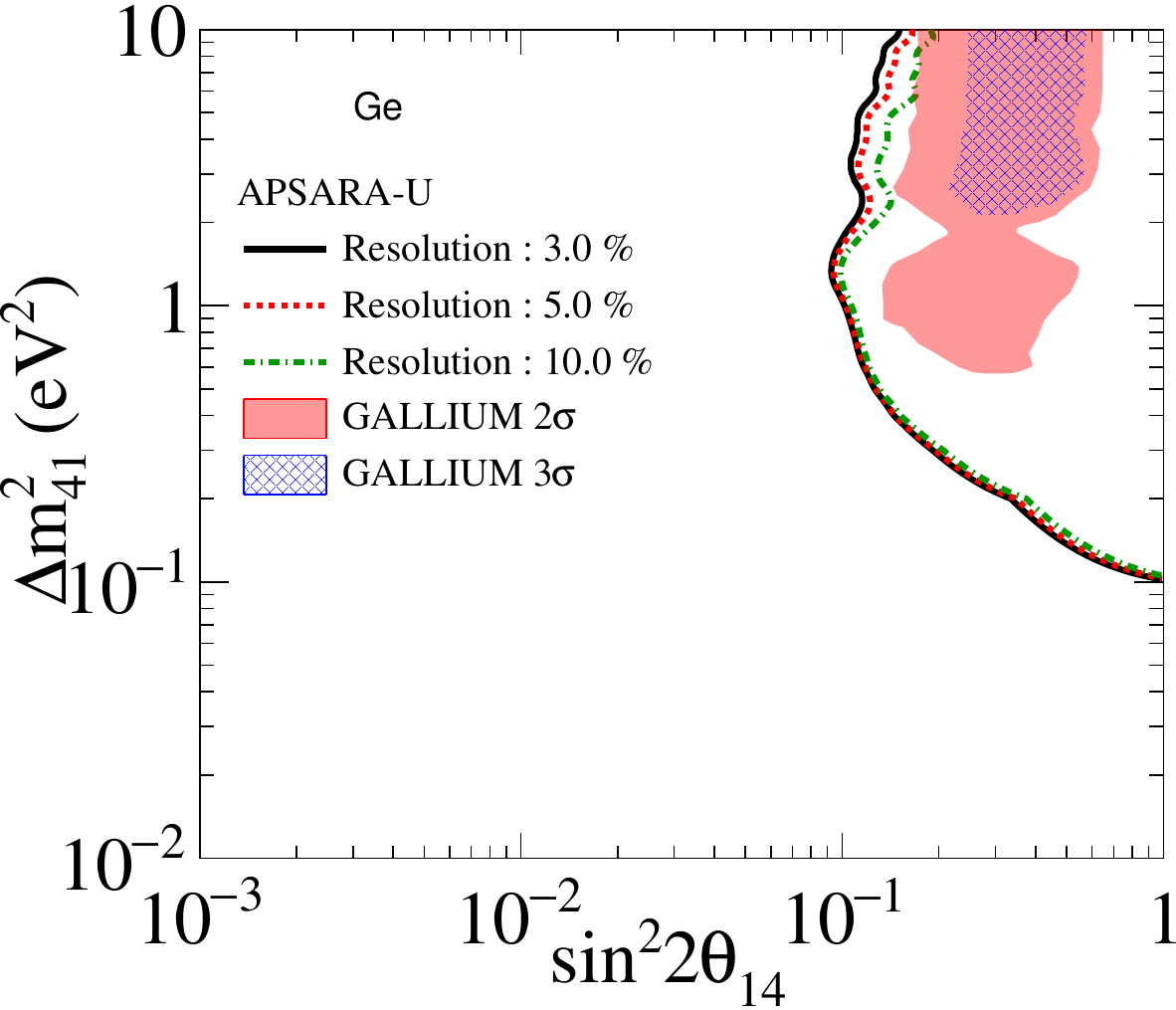}
\caption{ \label{fig:detReso} The comparison of ASN mixing sensitivity at 95 $\%$ C.L.
  for different resolution of Ge detector placed at 4 m from the U-Apsara reactor facility.
} 
\end{figure}

The sensitivity of the detector to the neutrino mixing parameters such as angle and 
squared mass difference depends on its resolution and efficiency. To investigate the
effect of energy resolution, we have considered a 10 kg Ge detector positioned 
at a distance of 4m from the U-Apsara reactor core. The detector resolution varies from 
$\sigma/E$ = 3\%--10$\%$/$\sqrt{E}$ for extracting its sensitivity to the upper limit 
for the ASN mixing angle $\theta_{14}$. Figure~\ref{fig:detReso} shows the 
sensitivity to ASN mixing parameters in the $\Delta m_{41}^2$ - $\sin^22\theta_{14}$ 
plane. It is observed that, the detector exhibits better sensitivity to the
ASN mixing with a resolution of $\sigma/E$ = 3\%$/\sqrt{E}$. Interestingly,
for $\Delta m_{41}^2 < 1.0$ eV$^2$, ASN mixing sensitivity remains unaffected by
  the detector resolution. However, for higher $\Delta m_{41}^2 \geq 1.0$ eV$^2$,
the detector's resolution improves the ASN mixing sensitivity.
At higher $\Delta m_{41}^2$, the oscillation 
frequency is more, and hence to have better sensitivity a detector with very good
energy resolution is necessary.
\begin{figure}
\centering
\includegraphics[width=0.45\textwidth]{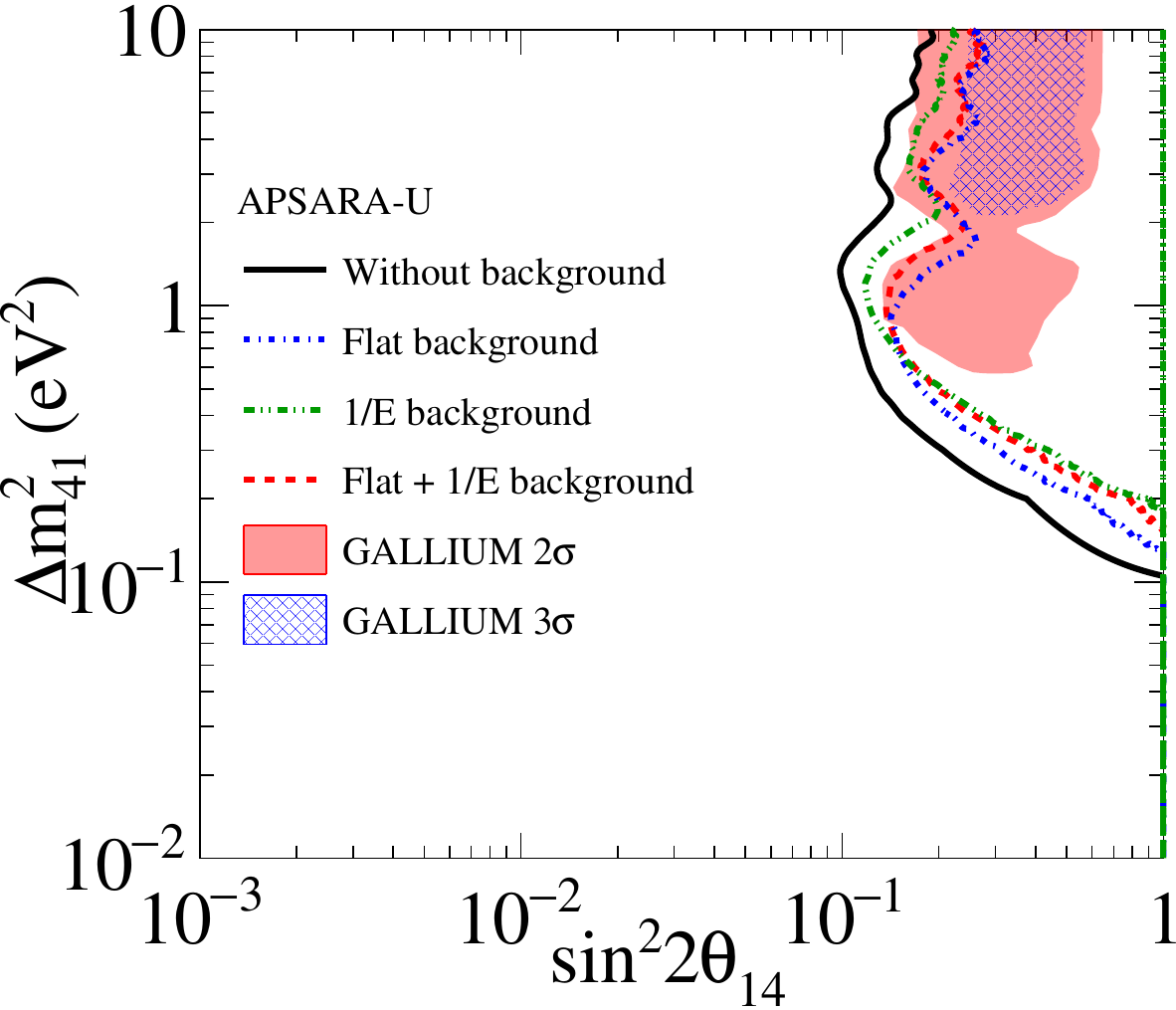}
\caption{ \label{fig:bkg} The comparison of ASN mixing sensitivity at 95 $\%$ C.L. of
  Ge detector in presence of background placed at 4m from the U-Apsara reactor facility.
  The signal to background ratio is one considered here.} 
\end{figure}
\subsection{Sensitivity of Detector in Presence of Background}
Measuring the low energy of recoil nuclei from the CE$\nu$NS process poses a
    significant challenge due to the presences of background. The background
  affects the detector sensitivity while extracting the various physics
  parameters of interest.  
The ASN mixing sensitivity has been obtained by incorporating backgrounds,
particularly in the context of Ge detector which will be placed in the
above-ground conditions at 4.0m distance from U-Apsara reactor.
At the experimental site, the reactor-related 
background (neutron and gamma) and cosmogenic background (muon-induced neutrons)
that can not be eliminated even with the shielding and 
will contaminate the actual signal. It is not only the types of backgrounds but also the
  energy dependent shape of the background that affects the detector sensitivity.
  At low recoil energy, 
there are two types of background shapes such as $1/T$ and flat-shaped backgrounds as
mentioned in Ref.~\cite{Bowen:2020unj}.
The simulated energy dependent background level due to neutron 
as well as gamma-rays are mentioned in Ref.~\cite{Strauss:2017cuu}. In this study,  
the energy dependent background shape is considered from Ref.~\cite{Bowen:2020unj} with signal
to background ratio of 1.0. While estimating the $\chi^2$, an 
associated 10$\%$ systematic uncertainty is therefore considered due to these 
backgrounds. Figure~\ref{fig:bkg} shows the comparison of Ge detector sensitivity 
with and without the inclusion of backgrounds. It is observed that with the 
contribution of both backgrounds, the ASN mixing angle sensitivity is further reduced 
as compared to the case with no background. Therefore better background reduction 
techniques have to be employed for the measurement with required sensitivity.
\section{Summary}
\label{sec:summary}
The study on neutrinos provides avenues for exploring numerous phenomena in 
physics beyond the standard model. At present, several experiments are underway
to measure various fundamental properties of neutrinos emanating from different sources. 
A proposed experimental program at the U-Apsara research reactor,
at BARC in India is described which aims to measure CE$\nu$NS. We have studied active-sterile
neutrino mixing sensitivity with Si, and Ge 
detectors in the context of CE$\nu$NS measurements at reactors with different core 
configurations and sizes available in India. The analysis has been carried out to
determine the active-sterile neutrino mixing sensitivity considering an exposure
of 1 year and detectors with varying masses, detection thresholds, and resolutions.
The region in 
sin$^2 2\theta_{14}$-$\Delta m^{2}_{41}$ plane is constrained by considering a single 
detector which will be placed at a fixed position with respect to the reactor core. 
It is found that the ASN oscillation at 95$\%$ confidence level with 
$\sin^{2}2\theta_{14}\geq 0.09$ at $\Delta m^{2}_{41}$ = 1.0 eV$^{2}$ can be observed 
with the Ge detector of mass 10 kg for an exposure of 1-yr. Additionally,
the Ge detector can exclude a significant portion of the favored non-zero
ASN mixing parameters region obtained from the GALLIUM experiment. Furthermore, the
sensitivity improves when placing the detector at PFBR or VVER reactor facilities. 

\bibliography{draft}
\bibliographystyle{apsrev4-1}

\end{document}